\documentstyle{article}
\begin{document}
\begin{center}
{\bf Possibilities of the Discrete Fourier Transform for Determining the Order-Chaos Transition in a Dynamical System}\\
{\it Carlos R. Fadragas(1), Ruben Orozco-Morales(2), Juan V. Lorenzo-Ginori(2),\\(1)Physics Department, Central University of Las Villas, Santa Clara, Cuba,\\(2)Centre for Studying Electronics and Information Technologies,\\ Central University of Las Villas, Santa Clara, Cuba.}
\end{center}
\begin{center}
{\bf Abstract}
\end{center}
This paper is devoted to a discussion of the Discrete Fourier Transform (DFT) representation of a chaotic
finite-duration sequence. This representation has the advantage that is
itself a finite-duration sequence corresponding to samples equally spaced in
the frequency domain. The Fast Fourier Transform (FFT) algoritm allows us an effective
computation, and it can be applied to a relatively short time series. DFT representation requirements were analized and applied for determining the order-chaos transition in a nonlinear system described by the equation $x[n+1]=rx[n](1-x[n])$. Its effectiveness was demonstrated by
comparing the results with those obtained by calculating the largest
Lyapounov exponent for the time series set, obtained from the logistic
equation.

\section{\bf Introduction}
One of the most studied problems in the scientific research deals with the processing of a time series $x[1],x[2],\ldots,x[N]$, composed by an experimental data sequence, spaced in time domain, sometimes evenly, sometimes unevenly. Such a sequence is obtained by sucessively sampling over a dynamical observable, characterizing a dynamical system under investigation. The goal is, as a result of processing, to obtain information about the behavior of the dynamical system that produced the time series. When we deal with a linear dynamical system, obtainnig  the information is relatively easy. However, there exist some difficulties when the dynamical system is a nonlinear one. In this case, the behavior of the dynamical system is rich enough. This was known for a long time, but it was not until the latest 20 years that the development of modern technologies and the digital numerical computation allowed finding solutions to important problems of the nonlinear dynamics, and to develop new methods in this field. The nonlinear dynamical analysis of a time series is a modern tool for scientific research, that has demonstrated its usefulness when it is to be applied in many branches of human knowledge. 

The analysis procedure begins by reconstructing a phase space portrait of the attractor by using delayed coordinates where the dynamics of the system can be observed\cite{Bleek,Fojt,Eckhardt,Gibson,Vibe}. For doing this, vectorial representations using delayed coordinates from a scalar time series ${x[n]};n=1,2,\ldots,N$, ($N$ being the length of experimental data sequence), are made up. The embedding dimension in the phase space, $D_e$, determines the dimensionality of the vectorial field $X[k]$
\begin{eqnarray}
X[k]=(x[k], x[k+\tau], x[k+2\tau], \ldots, x[k+(D_e-1)\tau])
\end{eqnarray}
It is usual that the time delay, $\tau$, is an integer number times the signal sampling period. The maximum number of state vectors that can be reconstructed with delayed coordinates, given a experimental data sequence of length $N$, equals $M=N-(D_e-1)\tau$. The vector set ${X[k]};k=1,2,\ldots,M$, each with coordinates  $x[k+i\tau];i=0,1,2,\ldots,D_e-1$, signals to points on the orbit of the system attractor in the phase space, so that the time evolution of the system in the phase space configuration can be observed by taking the initial condition $X[1]$ as a starting point and going on the behaviour of the vector state $X[k]$. This phase-space portrait is a representation of the behaviour of the dynamical system and can give us some information about the dynamics of the system\cite{Aradi,Fraser,Wolf,Fojt,Bleek,Eckhardt,Eckmann}. The selection of the values of reconstruction parameters, $D_e$ and $\tau$, can be made by using various methods\cite{Buzug,Albano,Fraser,Rosenstein,Yang,Blanco,Bleek,Takens,Grassberger,Farmer,Eckmann,Kennel,Fojt}.

After determining the reconstruction parameter values, one proceeds to calculate dimensions, exponents, and entropies. However, these quantities are difficult to compute, require a large data set, and degrade rapidly with additive noise. Generally, algorithms assume the repeatition of a same task for defining an adecuate value of the parameter that is being investigated\cite{Grassberger,Rosenstein}. Many authors recognize that the application of metric tools of the non linear dynamical analysis implies a high computational cost. It must be noticed that there are certain requirements related to the minimal length $N$ of the time series to be processed in order for the results obtained by applying nonlinear dynamical analysis to be realiable\cite{Eckmann}. 

Besides, according to the bibliography the transition from a regular to a chaotic behaviour has been studied in many dynamical systems, including the biological type ones. One of these is a fluid flow, where by calculating nonlinear dynamical parameters is possible to determine the order-chaos transition. CM Van den Bleek y JC Schouten\cite{Bleek} determined the order-chaos transition in a fluidized bed by calculating correlation dimension, $D_2$ and Kolmogorov entropy $K$ as a function of the Reynolds number and the superficial gas velocity, indicating two different regimes, order and chaos, and the intermediate region. 

On the other hand, the discrte-time Fourier transform (DFT) representation and the digital technique for its processing, known as fast Fourier transform (FFT) has been widely used for analysing a time sequence and a linear time-invariant system. However, it is required to distinguish three types of time series: deterministic, chaotic, and random. This classification is based upon the value that the Kolmogorov entropy takes in each case\cite{Bleek}. The DFT representation of a deterministic time series has been widely used, and the DFT representation of a random time series there doesn´t exist. We are interested now in analysing the possibility of applying the DFT representation to a chaotic finite-duration time series. We assume that the behaviour of a nonlinear dynamical system which produces a time sequence ${x[n]}; n=1,2,3,\ldots,N$, can change from a order to a chaotic regime and viceversa, and that is reflected in a any way in the time series observed\cite{Bleek}.

It is postulated in this work that every method, that allows us to give an evaluation about the behaviour of a dynamical system, and that can act over a relatively short time series, has a great practical importance. 

The aim of this research is to evaluate the ability of the DFT representation of a discrete-time series for obtainning some criteria about the ordered-chaotic transition related to the behaviour of a dynamical system, without applying {\it a priori} discussed and laborious nonlinear dynamical analysis algorithms, characterized all of them by a high computational cost. In particular, the method proposed in this paper will be realiable if it is applied to discrete-time series set generated by a given dynamical system, where the comparative criteria are to be the goal. It is important to say that this method does not pretend to sustitute those of the nonlinear dynamical analysis, but it can be used for analysis complementation. So that, the DFT representation can be applied to a shorter time series than those required by the nonlinear dynamical analysing.

We produced a 120 time series set with 1024 data points in each time series, and for each of them we obtained its 1024 FFT representation. By observing the amplitude spectrum and the phase spectrum behaviour with the control parameter of the system, a well defined value of the control parameter ($r\cong 3.57$) can be determined, beyong which the behaviour either the amplitude spectrum and the phase spectrum became quite irregularly, in contrast with the regular behaviour of these magnitudes in the region before the indicated value above. There exists a complete correpondence when this result is compared with that obtained from the bifurcation diagram and from the largest Lyapounov exponent graphic, both as a function of the control parameter $r$.

\section{\bf Methods and Materials}

\subsection{\bf Obtention of the discrete-time series set}

The logistic equation is a one-dimensional mapping of the real axis in the interval (0,1), that is a prototype of a nonlinear dynamical system widely used\cite{Cohen} and can be formulate as
\begin{eqnarray}
x[n+1]=rx[n](1-x[n]),
\end{eqnarray}
and with the initial condition $x[1]$ especified, one can obtained a time sequence of a given length by a recursive action. The character of the behaviour of a data sequence obtained with the iterative process can be modified by selecting conveniently a value of the control parameter $r$. In this work a value of the initial condition $x[1]=0.65$ was taken, and the interval $r \in [2.8,4.0]$ was selected. A set of evenly spaced values of $r$, with a step $\delta=0.01$, was taken to produce a family of 120 discrete-time series, each containning $N=2^{10}$ samples. The building up of the histogram for each time series allows us to give inmediately a simple statistical characterization of these time series. 

In order to do much easier further computations data points were organized into a rectangular matrix containing 120 columns each of which is a time series with the length $N=1024$ data points. 

\subsection{\bf Determining of the largest Lyapounov exponent $\lambda_1$}
Lyapounov exponents spectrum deals with average exponential divergence or convergence of two neighboor orbits in the phase space. It is common to order the spectrum from largest to shortest exponents: $\lambda_1,\lambda_2,\lambda_3,\ldots$. A system containning one or more possitive exponents is to be defined as chaotic. For calculating the Lyapounov exponents stectrum some algorithms have been developed, according to the particular situation, and being one of the most used algorithm that reported by Wolf et al\cite{Wolf}. From the exponents spectrum, the largest exponent, $\lambda_{1}$, decides the behaviour of the dynamical system. For calculating the largest exponent one can refer too the algorithm proposed by Rosenstein et al\cite{Rosenstein}. For estimating $\lambda_1$ one can monitor the large term evolution of two neighboor orbits on the attractor reconstructed. Details of these algorithm can be found in refered papers.  

We determined the largest Lyapounov exponent,$\lambda_1$, for each time series in the set, using a professional software\cite{Sprott}. The selection of the time delay, $\tau$, and the embedding dimension, $D_e$, required for reconstructing  the phase space is part of the problem and it is necessary to consider some selecting criteria. This selection can be made by different methods\cite{Bleek,Takens,Grassberger,Farmer,Eckmann,Kennel,Fojt}. For selecting $\tau$ one can cite:(a) Optimal filling of the phase space\cite{Buzug}, (b) by determining the position of the first local minimum of the data autocorrelation function\cite{Albano}, (c) by taking the first minimum in the mean mutual information plot\cite{Fraser}, (d) based on the optimal tradeoff between redundance and the irrelevance\cite{Rosenstein}, and by analysing of the differential equations of the system if they are given\cite{Yang}. For selecting $D_e$ can be indicated: (a) by applying the Grassberger-Procaccia method, which allows one to obtain simultaneously both correlation dimension and embedding dimension\cite{Grassberger}, (b) by applying the false nearest neighboors method\cite{Rosenstein,Blanco}, and (c) by applying geometric considerations\cite{Kennel}.

A time series from the logistic equation was analyzed by using the method (a) for $\tau$ determining and by taking a value $D_e=3$, which satisfies the Grassberger-Procaccia algoritm condition $D_e>2D_2$, and Takens theory\cite{Grassberger,Takens}, being $D_2=0.5$ a value reported by Hoyer et al\cite{Hoyer} for this time series type. Figure 4 shows the effect of optimal filling of the phase space for one of the time series obtained here.

Literature refers\cite{Grassberger,Takens} that selecting embedding dimension as
\begin{eqnarray}
D_e \geq D_2+1,
\end{eqnarray}
is suffient to have a good representation for the attractor in the phase space, reconstructed in delayed coordinates from a scalar time series observed. Any reconstruction made with a value of the embedding dimension below the minimal value will produce a projection, in that dimension, of the original attractor, and therefore it will be more difficult for the interpretation of that projection or it will produce false results when a estimation of nonlinear dynamical parameter is to be made. It is important to take a value for embedding dimension close to $2D_2+1$ because of it is not the case, computational calculation will be excesively made.

\subsection{\bf Application of DFT representation: Amplitude spectrum and phase spectrum.}
The DFT representation of a finite-length discrete-time series $x[n];n=0,1,2,\ldots,N-1$, is given by
\begin{eqnarray}
x[n]=\frac{1}{N}\sum_{k=0}^{N-1}X[k]e^{j(2\pi/N)kn};n=0,1,2,\ldots,N-1\\
X[k]=\sum_{n=-0}^{N-1}x[n]e^{-j(2\pi/N)kn};k=0,1,2,\ldots,N-1
\end{eqnarray}
The sequence $X[k]; k=0,1,2,\ldots,N-1$ can be efficiently computed by a digital algorithm known as Fast Fourier Transform (FFT), from which the amplitude spectrum, $\vert X[k]\vert$ or ${\it abs}(X[k])$, and the phase spectrum, ${\it arg}(X[k])$ or ${\it angle}(X[k])$, are both obtained.

Next we obtain the DFT representation for each time series in the set by the 1024 points-FFT of the set matrix. Since each time series produced by the logistic equation (with $r \in [2.8,4.0]$) is a finite-duration bounded sequence, it satisfies the existence requirement for the DFT representation, that is
\begin{eqnarray}
\sum_{n=-\infty}^{\infty}\vert x[n] \vert< \infty
\end{eqnarray}
We considered that it is convenient, for some purposes, to eliminate the dc component of signals produced by the logistic equation, before the amplitude spectrum mesh plot is made.

\subsection{\bf Results}
Figures, from 1 to 3, show a sample of three representative time series and their corresponding histograms. In order to give a better view of each time series it was only considered 128 data points of each discrete-time series in its plotting but it was completely considered when its corresponding histogram was made. In either case, the particular value of $r$ parameter was indicated.

The two-dimensional plot of the rectangular matrix produces a figure known as Feigenbaum or bifurcation diagram. It reflects the transition of a dynamical system towards chaos by a period-doubling process. Feigenbaum diagram are obtained when parameter values are plotted on the horizontal axis and sequence data values are plotted in the vertical axis (see appendix). There exists a threshold value in $r\cong 3.57$ above which the transition of the dynamical system to the chaotic behavior can be observed. The bifurcation diagram serves as a reference to control results, and it is usually applied in this research frame\cite{Aradi,Rodelsperger}.

Figure 5 shows results for the largest Lyapounov exponent estimate $\lambda_1$, calculated for each of the time series in the set, as a function of the control parameter $r$. This reference is taken as a control for the discussion of the results obtained by applying the DFT representation, and it is not a particular aim of this work. 

Figure 6 shows a mesh plot of the DFT representation amplitude spectrum $abs(X[k])$, (in order to obtain a finest version of each figure and figure showing a mesh plot of the DFT representation phase spectrum $angle(X[k])$, please, see appendix),for the discrete-time series matrix. The {\it x} axis corresponds to the control parameter $r$, and the {\it y} axis corresponds to the frequency, from zero value to the Nyquist frequency value. In order to obtain the best view of the DFT representation amplitude response mesh plot, the dc component on the signal was eliminated.

\section{\bf Discussion of the results}
The application of the DFT representation of a chaotic-type signal gives satisfactory results for determining the order-chaos transition in the logistic equation system. This conclusion is confirmed by comparing that plot given in Figure 6 with that shown in Figure 5, where the behavior of the largest Lyapounov exponent $\lambda_1$ with the control parameter $r$ is shown. The largest Lyapounov exponent is one of the metric tools used for evaluating the system dynamics. Besides, bifurcation diagram corresponding to the logistic equation system may be used too as a reference. A critical value for control parameter $r$ is reported in literature above which sequences produced by the logistic equation exhibit, all of them, a chaotic behavior. This threshold value is around $r\cong 3.57$, as is shown in Figure 5.

Some intermediate ordered region in the chaotic behavioral region underlie, but some authors\cite{Wolf} refer them as a consequence of the computational system limitations rather than a real behavior of a dynamical system. However, this relationship between order and chaos can be observed from the dependence of the largest Lyapounov exponent $\lambda_1$ with control parameter, $r$, whenever the largest Lyapounov exponent value tends to zero at bifurcation points embedded in chaotic region ($r\geq 3.57$). The behavior of a dynamical system, for  control parameter values correponding to largest Lyapounov exponent close to zero, is not chaotic. 

The reconstruction parameter values selected are in correpondence with literature, and satisfy requirements of the Takens theory\cite{Takens}, and Grassberger and Procaccia algoritm\cite{Grassberger}. The bifurcation diagram was builded up several times, changing the initial condition, and changing the step value for the control parameter, and the feature of the bifurcation diagram for $r\geq 3.57$ remains. It can be remarked that the order-chaos transition for the value $r\cong 3.57$ can be observed in DFT representation amplitude spectrum mesh plot (and in DFT representation phase spectrum mesh plot, see appendix), shown in Figure 6.

On the other hand, Figures from 1 to 3 show the time domain representation of some of time series and their corresponding histogram. Comparing Figure 2 ($r=3.70$) and 3 ($r=3.83$), it can be deduced the bifurcation effect from their histogram. Note that for figure no.3 the control parameter satisfies $r>3.57$, and it corresponds to a value for the largest Lyapounov exponent close to zero for $r\cong 3.83$, as it can be estimated in Figure 5. 
It can be remarked that the bifurcation diagram analysis is used in studies of the behavior of a dynamical system that tends to chaos through period-doubling procedure\cite{Aradi}.

Applying DFT representation to discrete-time series in a family derived from a observable in a dynamical system is recommended for qualitative detecting of order-chaos transition, and this can be add to the metric tools of the nonlinear dynamical analysis. Any method that allow us to evaluate the behavior of a dynamical system, and that can be applied to a relatively short discrete-time series, must be taken into account as a practical method.

\section{Conclusions}
The application of the time series DFT representation can be made over a chaotic-type time series, nevertheless this time series is to be a nonlinear one. This method have not strong requirement about the length of the experimental data sequence. If one deals with a nonlinear dynamical system that can modify its behaviour between order and chaos, a time series produced by the system reflects this change, and the DFT representation of that time series exhibits in its amplitude response and its phase response the change in the system too. Any method, that allow us to give some evaluation about the behaviour of a dynamical system and that can work over a time series of a short length, adquires a practical importance. The DFT representation can be used for complementing the results obtained by applying the most powerfull methods of the nonlinear dynamical analyisis.
\newpage
\section{Appendix}
In order to see the bifurcation diagram, a finest version of all figures including the DFT representation phase spectrum mesh plot, the following program may be runned on MatLab. For running on MatLab, write a percent symbol at the begining of each comment line. Comment: [Figures for DFTQI paper.] Comment: [To producing a family consisting of 120 time series, each of one having 1024 data points.] close all, clear, Comment: [Datos, $r=2.8$; $del=0.01$; $nr=120$; $x=0.65$; $N=1024$;] for $j=1:nr$;$R(j)=r$;$r=r+del$;end; for $j=1:nr$;$for i=1:N$;$A(i,j)=x$;$x=R(j)*x*(1-x)$;end;end; Comment: [Figures selected, $j=[14\;18\;21\;24\;61\;72\;91\;104\;109]$;] for $k=1:9$;$rf(k)=R(j(k))$;end;rf, disp('values of j equals to 14, 94, y 104 were selected'); $R=[2.93\;3.70\;3.83]$; pause(1); figure(1); subplot(211); plot(A(1:128,14)); title('Figure 1:Time series and its histogram, for $r=2.93$'); xlabel('Time index'); ylabel('Sequence values'); subplot(212); hist(A(1:128,14),40); xlabel('Sequence values'); ylabel('Relative frequency'); pause(1); figure(2); subplot(211); plot(A(1:128,94)); title('Figure 2: Time series and its histogram, for $r=3.70$'); xlabel('Time index'); ylabel('Sequence values'); subplot(212); hist(A(1:128,94),40); xlabel('Sequence values'); ylabel('Relative frequency'); pause(1); figure(3); subplot(211); plot(A(1:128,104)); title('Figure 3: Time series and its histogram, for $r=3.83$'); xlabel('Time index'); ylabel('Sequence values'); subplot(212); hist(A(1:128,104),40); xlabel('Sequence values'); ylabel('Relative frequency'); Comment: [For optimal filling criterium.] $r=2.8$; $nx=2^{10}$; $del=0.01$; $nr=120$; for $j=1:nr$; $R(j)=r$; $x0=0.65$; $v0=0.01$; $x=x0$; $v=v0$; for $i=1:nx$; $A1(i,j)=x$; $B1(i,j)=v$; $z=x$; $x=r*x*(1-x)$; $v=x-z$; end; $r=r+del$; end; pause(1); figure(4); subplot(221); plot(A1(1:nx,114),B1(1:nx,114)); title('Figure 4a : $r=3.94$'); 
Comment: [Part of the program for largest Lyapounov exponent ploting.] $xr=2.82:0.02:2.8+60*.02$; 
\begin{eqnarray}
yexl=[-0.572\; -0.240\; -0.220\; -0.180\; -0.153\; -0.118\; -0.091...\nonumber\\
   -0.052\; -0.031\; -0.000\; -0.055\; -0.127\; -0.201\; -0.284\; -0.381...\nonumber\\
   -0.492\; -0.625\; -0.781\; -0.996\; -1.310\; -1.250\; -0.855\; -1.030...\nonumber\\
   -1.166\; -0.886\; -0.686\; -0.536\; -0.407\; -0.295\; -0.198\; -0.112...\nonumber\\
   -0.034\; -0.080\; -0.340\; -1.445\; -0.280\; -0.036\; -0.111\; 0.148...\nonumber\\
   0.258\; 0.288\; 0.320\; 0.406\; 0.497\; 0.526\; 0.524\; -0.120...\nonumber\\
   0.577\; 0.589\; 0.651\; 0.608\; -0.067\; 0.478\; 0.661\; 0.710...\nonumber\\
   0.752\; 0.797\; 0.825\; 0.850\; 0.945]\nonumber
\end{eqnarray}
pause(1); figure(5); plot(xr,yexl,'kx'); hold on; plot(xr,yexl,'k:'); grid on; xlabel('Control parameter'); ylabel('Largest Lyapounov exponent'); title('Figure 5: Behavior of LLE with r'); legend('Experimental values','Behavior',0);zoom on; Comment:Part of the program for calculating and ploting the amplitude spectrum mesh and the phase spectrum mesh of the DFT representation. 
$A=A-ones(N,1)*mean(A)$; $f=linspace(0,pi,N/2)$; $ftA=fft(A)$; $amftA=abs(ftA(1:N/2,:))$; $anftA=angle(ftA(1:N/2,:))$; pause(1); figure(6); mesh(R,f,amftA); title('Figure 6: Amplitude spectrum mesh'); xlabel('Control parameter'); ylabel('Frequency'); zlabel('Amplitude'); view(-38,22); pause(1); figure(7); mesh(R,f,anftA); title('Figure 7: Phase spectrum mesh');xlabel('Control parameter'); ylabel('Frequency');zlabel('Phase');view(-40,30);

\newpage
\begin{figure}[b]
\caption{Time series and its histogram, for r=2.93}
\end{figure}
\begin{figure}[b]
\caption{Time series and its histogram, for r=3.70}
\end{figure}
\begin{figure}[b]
\caption{Time series and its histogram, for r=3.83}
\end{figure}
\begin{figure}[b]
\caption{Optimal filling of the phase space criterium}
\end{figure}
\begin{figure}[b]
\caption{Behavior of the largest Lyapounov exponent}
\end{figure}
\begin{figure}[b]
\caption{Amplitude spectrum mesh}
\end{figure}
\end{document}